# Single-cycle, MHz-repetition rate THz source with 66 mW of average power

F. MEYER*, T. VOGEL, S. AHMED, AND C.J. SARACENO

*Photonics and Ultrafast Laser Science, Ruhr Universität Bochum, Universitätsstrasse 150, 44801 Bochum, Germany*
*Corresponding author: Frank.Meyer-h1j@Ruhr-Uni-Bochum.de*

**We demonstrate THz generation using the tilted pulse front method in Lithium Niobate, driven at unprecedented high average power of more than 100 W and at 13.3 MHz repetition rate, provided by a compact amplifier-free modelocked thin-disk oscillator. The conversion efficiency was optimized with respect to pump spot size and pump pulse duration, enabling us to generate a maximum THz average power of 66 mW, which is the highest reported to date from a laser-driven, few-cycle THz source. Furthermore, we identify beam walk-off as the main obstacle that currently limits the conversion efficiency in this excitation regime (with moderate pulse energies and small spot sizes). Further upscaling to the watt level and beyond is within reach, paving the way for linear and nonlinear high-average power THz spectroscopy experiments with exceptional signal-to-noise ratio at MHz repetition rates.**

Terahertz time domain spectroscopy (THz-TDS) has been one of the cornerstones of THz science for several decades, leading to a vast number of advances in science and technology. One example in scientific research where THz-TDS continues to enable major breakthroughs is in condensed matter physics, where time-resolved spectroscopic techniques (pump-probe, multi-dimensional spectroscopy) in the THz region have become ubiquitous [1]. However, in this and many other fields, new challenges keep arising, as THz-TDS is performed to study increasingly difficult samples, such as water with large THz absorption or at long distances, such as in imaging applications and/or sensing. These and other applications would benefit from greater signal-to-noise ratios and shorter measurement times, enabled by high repetition rates, while maintaining sufficient THz pulse energy – i.e. from sources with high average power. Experiments requiring high average powers are thus typically exclusively performed in accelerator facilities, at the expense of extremely high cost, very restrictive accessibility and difficulties in achieving phase stability required for TDS [2]. Meanwhile, lab-based few-cycle THz sources for TDS have immensely progressed in terms of pulse energy [3,4], bandwidth [5] and tunability [6] in recent years, but this progress is typically made at the expense of the repetition rate of the source (<1 kHz), therefore average power levels remain very low, typically in the few to hundreds of μW range. This is mostly due to the limited average power of commonly used energetic Ti:Sapphire amplifiers as drivers, which are typically limited to <10 W of average power.

One promising yet widely unexplored path to increase the average power of current THz sources is therefore to adopt novel ultrafast solid-state lasers based on Yb-doped gain media as driving sources, which nowadays reach kilowatt average powers [7–9], possibly in combination with efficient THz generation schemes. Among these novel ultrafast technologies, we focus our attention on modelocked thin-disk lasers (TDLs), which are capable of providing femtosecond pulses with tens of microjoules of pulse energy and hundreds of watts of average power directly from a modelocked oscillator operating at multi-MHz repetition rate [10]. In spite of the potential of these, as well as other high-average power amplifier systems, only few attempts have been made so far to drive THz generation with the available state-of-the-art high driving average powers, most likely because of the large difficulties related to heat handling and damage in most commonly used media for frequency down-conversion.

Very recently, THz generation in two-color plasma filaments was explored in combination with high-average power excitation because of its intrinsically damage-free operation and pump pulse duration limited THz bandwidths, with conversion efficiencies in the $10^{-4}$ range. In this experiment, a 120 W Yb-fiber amplifier system was used to generate 50 mW at a repetition rate of 100 kHz [11], which represents the highest THz average power demonstrated so far. However, in this result the THz waveform and spectrum have not been fully characterized yet, limiting the application possibilities of this system. Furthermore, much higher repetition rates in the MHz regime are still challenging, due to the requirement of very high driving pulse energy in the mJ range.

Optical rectification (OR) in $\chi^{(2)}$ crystals is a promising path to make use of high average power driving lasers, in particular for systems with MHz repetition rates. Most recently, we demonstrated a first result in this direction using collinear OR in GaP, driven by a 112 W modelocked thin-disk oscillator, nonlinearly compressed to 88 fs at 13.4 MHz repetition rate [12], reaching 1.35 mW of THz average power. However, in this experiment we were limited to moderate conversion efficiencies (~$10^{-5}$) mostly due to well-known multi-photon absorption effects. In this respect, OR in Lithium Niobate (LN) is better suited to reach high average power levels due to its large effective nonlinearity and because it can withstand much higher intensities before multi-photon absorption effects are dominant. However, in this technique, the pulse front of the driving laser needs to be tilted in order to achieve phase matching over significant propagation lengths [13], thus significantly increasing complexity, and the THz bandwidth is limited to <4 THz due to phonon absorption [14]. Using this scheme, a record conversion efficiency of 3.7% has been achieved with cryogenically cooled LN. This resulted in 44 mW of THz power at a moderate pump power of 1.2 W and 1 kHz repetition rate [15]. To the best of our knowledge, only one attempt has been made in this geometry using MHz repetition rate driving lasers, where an Yb-fiber amplifier with 14 W at 1 MHz was used to achieve 250 µW THz average power [16].

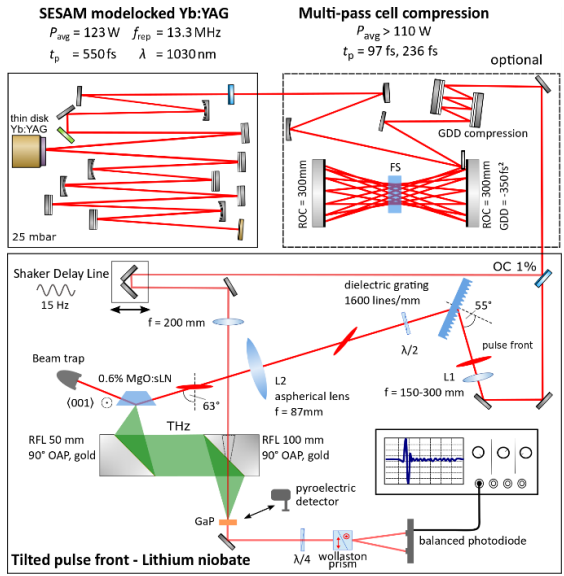

Fig. 1. Experimental setup consisting of modelocked thin-disk oscillator, optional MPC as well as the THz generation and detection setup.

In this letter, we demonstrate OR in the TPF geometry in LN at room temperature, driven by a modelocked thin-disk oscillator with >110 W average power at 13.3 MHz repetition rate. Optimization of both spot size and pulse duration of the driving laser results in 66 mW THz average power, which to the best of our knowledge is the highest so far achieved from a few-cycle ultrafast laser-driven THz source. Furthermore, this is the highest repetition rate and pump average power ever used with this technique, showing its potential for average power scaling. We identify several obstacles that are unique to this new excitation regime and discuss possibilities for further upscaling.

The full experimental setup is shown in Fig. 1a. The driving laser was a home-built semiconductor saturable absorber mirror (SESAM) modelocked Yb:LuAG thin-disk oscillator delivering 123 W at 13.3 MHz repetition rate with a central wavelength of 1030 nm and a pulse duration of 550 fs, resulting in 9.2 µJ pulse energy. In order to have different pulse durations available, a Herriott-type multi-pass cell (MPC) was used to reduce the pulse duration via spectral-broadening due to self-phase modulation and chirp compensation [17,18]. In our experiment, the compressor consisted of two concave mirrors with a radius of curvature of 300 mm placed at a distance of 540 mm. In the first configuration, the beam underwent spectral broadening in 42 passes through a 12 mm thick piece of AR coated fused silica (FS), and the pulse was subsequently compressed nearly to its transform limit by 24 bounces on two pairs of dispersive mirrors with -550 fs$^2$ group delay dispersion, reaching 97 fs pulses. The laser system and compressor are described in more detail in [12]. Compared to [12], we additionally realized an intermediate configuration consisting of 18 passes through a 6 mm thick FS plate and 40 bounces on the dispersive mirrors resulting in 236 fs pulse duration. This allowed us to perform the THz generation experiment with three different pulse durations (97 fs, 236 fs and 550 fs, see Fig. 2b, inset). The compressed pulses were characterized using a home-built second harmonic generation frequency-resolved gating (SHG-FROG) setup, while the sech$^2$-shaped pulses out of the oscillator were measured by an intensity autocorrelation. In both MPC configurations, we obtained excellent power transmission of the compressor, making >110 W available for the experiment with variable pulse duration.

We generated THz radiation by OR in a 0.6% MgO-doped stoichiometric LN trapezoid using a pulse front tilt of 63° to achieve phase matching between pump and THz fields. The tilt is generated using a dielectric transmission grating with 1600 lines/mm at an angle of incidence of 55.5° in combination with a 2" aspherical lens (L2) placed at a distance of 230 mm, which images the grating into the LN crystal with a focal length of 87 mm. This configuration was adopted according to the guidelines in [19]. A half-wave plate placed after the grating ensured vertical polarization. The pump beam was reflected off the front surface of the LN trapezoid by total internal reflection and subsequently exited the crystal, where it was dumped in a beam trap. Both the beam trap and the crystal mount were water-cooled. The THz beam exited the front surface of the trapezoid and was collected and refocused by two 2" off-axis parabolic mirrors (OAPs) with an effective focal length of 50 mm and 100 mm, leading to a magnification of M=2 for the THz spot. We characterized the THz radiation using either a pyroelectric power meter or a standard electro-optic sampling (EOS) setup, using a <1% fraction of the pump beam in a 1 mm GaP crystal, followed by a quarter-wave plate, a Wollaston prism and a balanced photo diode (PD). The delay was generated by a shaker delay line oscillating at 15 Hz, allowing us to display the PD signal directly on an oscilloscope and acquire 15 traces/s. In order to optimize the conversion efficiency, all crucial elements (grating, imaging lens, crystal) were placed on linear translation stages in order to fine-tune their relative position. Additionally, the grating was placed on a rotation mount for adjustments of the tilt angle.

The peak intensity inside the LN trapezoid was set by adjusting the spot size of the pump on the grating using either a f=150 mm, f=250 mm or f=300 mm lens (L1) at variable distances before the grating, which generated a focus behind it. The beam size inside the crystal was smaller by a demagnification factor M=0.66, caused by the imaging lens L2. Note, that placing the grating outside the focus of L1 lead to a finite wave front radius of curvature (ROC). The effect of this ROC will be discussed below. The whole setup could be purged with dry Nitrogen to less than 3% rel. humidity in order to reduce water vapor absorption in the air.

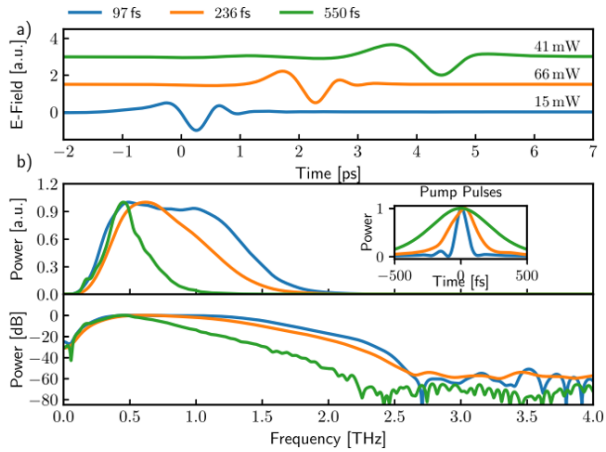

Fig. 2. EOS Results for three different pump pulse durations at full pump power. a) Electric Fields measured by EOS time domain. The traces were offset in both x- and y-direction for better visibility. b) Normalized power spectra plotted on linear and logarithmic scales. The inset shows the measured pump pulse profiles.

Fig. 2a shows the THz transients measured by EOS in the time domain for all three available pump pulse durations and at the full available power (see Fig. 3) under purged conditions. The setup was aligned for maximum EOS signal in all cases. Fig. 2b shows the corresponding normalized power spectra on linear and logarithmic scales. As expected, the bandwidth increases with decreasing pulse duration. The spectra are centered in the vicinity of ~0.7 THz as a result of phase matching and strongly increasing THz absorption towards the phonon resonance at 4 THz. The bandwidth extends to beyond 2 THz for all three pulse durations. We averaged over 20 traces, resulting in a SNR of ~60 dB in all cases. Note however, that the detection has not been optimized yet for highest SNR, but the focus of this experiment was on optimizing the generation setup.

We measured the THz power vs. pump power (before grating with 95% efficiency) for different beam diameters and pulse durations. Fig. 3a shows the measured slopes at 550 fs, for three different beam diameters on the grating. The highest THz power obtained under unpurged conditions was 41 mW for the 1 mm beam at 121 W pump power with a peak intensity of ~8.6 GW/cm$^2$ inside the crystal. When reducing the spot size to 0.7 mm (double intensity), we unexpectedly observed a reduction in THz power, which contradicts the trend observed in most other experimental realizations. In fact, typical peak intensities reported in the literature are on the order of 100 GW/cm$^2$ [3]. In order to avoid damage, we did not use the full pump power in this case.

Another way to increase the peak intensity is by using shorter pulses. THz powers obtained for different pulse durations under optimized focusing conditions along with the corresponding conversion efficiencies are shown in Fig. 3b. Using 97 fs pulses we obtained a lower conversion efficiency than with 550 fs pump pulse duration, whereas higher efficiency was obtained with the 236 fs pulse. In the optimal case (236 fs), we measured 66 mW average power at the maximum input power of 118 W, which is to the best of our knowledge the highest average power so far demonstrated from a single-cycle THz source. Taking into account the pulse shape from the EOS and the measured THz beam profile (see Fig. 4 below) we estimate a peak electric field of 16.7 kV/cm. This value can be further increased by choosing THz imaging optics with a magnification below the current value of M=2. The optimal pump spot size in this case was 1.4 mm, corresponding to approximately the same peak intensity of ~8.6 GW/cm$^2$ than the 550 fs case. No damage of the crystal was observed for any of the parameters reported here. The power values and EOS traces were stable and reproducible over the full measurement time span (up to 20 min), indicating no long-term thermal effects were present.

It is critical to notice here, that both the peak intensity and optimal conversion efficiency (<6·10$^{-4}$) are lower than typical values used/obtained with the TPF method with larger pump pulse energies, for example in the room temperature experiments reported in [3,15]. Using much smaller pulse durations and spot sizes and consequently higher peak intensities lead to a reduction in conversion efficiency. For shorter pulse durations, an increase of linear absorption for higher THz frequencies and a stronger impact of group velocity dispersion due to angular dispersion (GVD-AD) - which is a result of the pulse front tilt (PFT) and leads to a decrease in interaction length - are expected. In fact, pulse durations around ~500 fs were shown to be optimal in the TPF geometry with LN [20], contrary to what we observe here. However, in most studies performed so far, the influence of the beam diameter is usually not taken into account and has not been studied in depth, because typically mJ pump pulse energies with beam diameters of several mm or more are used. With sub-mm diameters, as used here, walk-off between the pump and THz fields due to the non-collinear nature of the experiment is non negligible. This is reflected by the strongly elliptical THz beam profile measured for the 66 mW THz beam with a microbolometer camera, as shown in Fig. 4a. Fig. 4b shows lineouts along the x-direction (horizontal, plane of PFT) and y-direction (vertical) together with gaussian fits. The resulting diameters are $x_0$=2.1 mm and $y_0$=0.6 mm. The magnification of M=2 by the OAPs is already taken into account, therefore these values represent the THz beam size at the LN crystal surface. Along the y-direction the THz beam size is mainly determined by the pump beam size, while along the x-direction the beam size is determined by the effective interaction length between pump and THz fields. Since $x_0$ is bigger than the used pump beam diameter, walk-off is significant here. The reduction in overlap between pump and THz fields then leads to a reduction in conversion efficiency, which explains the trade-off between larger beam sizes and peak intensity that we observe.

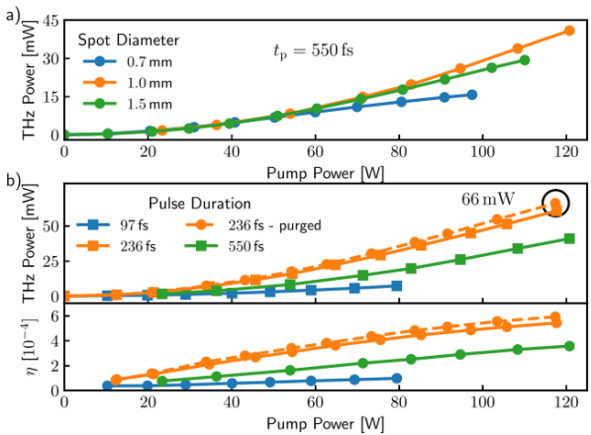

Fig. 3. a) THz power vs. pump power at 550 fs pulse duration for different spot sizes on the grating. b) THz power vs. pump power for three different pump pulse durations. The lower panel shows the corresponding conversion efficiencies η.

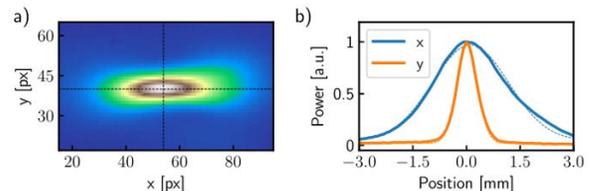

Fig. 4. a) THz beam profile for 236 fs pump at 66 mW average power. b) Lineouts of the THz beam along the axis indicated in a). The dashed lines represent Gaussian fits.

The reduction in efficiency with beam size and pulse duration has also been observed in numerical studies [21,22] showing qualitative agreement with our results. However, for a quantitative agreement, detailed numerical simulations are needed that take into account all spatio-temporal effects as well as pump depletion, which has been shown to limit the interaction length further [23].

Another contribution that needs to be taken into account is the finite ROC of the wave front, which leads to a change in the PFT along the propagation direction [21] and can therefore influence the efficiency by reducing the effective interaction length. The ROC inside the LN crystal in our specific setup was calculated using an ABCD-matrix formalism resulting in 0.05 m - 0.1 m for the spot sizes used here. The corresponding change in PFT angle for these ROCs over a propagation distance of 2 mm is 0.5° - 1°, which is significant. However, no significant change in the THz spot size along the x-direction is measured for any combination of spot size, ROC or pulse duration, which shows that the effective interaction length is not drastically influenced by the ROC. We therefore conclude that the walk-off effect is the main limiting factor in this experiment. Nonetheless, further improvement can be expected avoiding small ROCs by using longer focal lengths or employing a telescope instead of a single imaging lens.

While the walk-off effect seems to be the main limitation, other contributions might play a role as well: Thermal- or refractive effects due to the very large average power cannot be ruled out at this point. Furthermore, self-focusing will be more pronounced for small beam diameters than for larger ones and can contribute to pulse break-up and a limitation in interaction length. An indication, that these effects need to be considered can be found in the efficiency curves displayed in Fig. 3b, which show saturation for high pump powers/energies. This cannot be explained with walk-off or ROC alone, which do not depend on pump power. Further investigations are underway to disentangle these effects.

Nevertheless, one clear trend from our experiment emerges: higher pulse energies and consequently larger spot sizes should relax all the aforementioned issues. Thin-disk oscillators with pulse energies up to 80 µJ have been demonstrated [24], which is ~9x higher, than the pulse energies used here. This would allow to increase the used beam diameters by a factor of 3 at our current peak intensities, thereby significantly reducing walk-off. Further improvements can be expected by cryogenic cooling of the LN crystal, which has been shown to increase the conversion efficiency by a factor of 3.2 [15]. THz sources with average powers reaching the watt-level can therefore be expected in the very near future.

**Funding.** Deutsche Forschungsgemeinschaft (DFG) TRR196 – C10, und im Rahmen der Exzellenzstrategie des Bundes und der Länder – EXC 2033 – Projektnummer 390677874 – RESOLV. Alexander von Humboldt Foundation (Sofja Kovalevskaja Award endowed by the Federal Ministry of Education and Research).

**Disclosures.** The authors declare no conflicts of interest.